\documentclass[12pt]{article}
\usepackage{graphics}
\usepackage{epsfig}
\usepackage{hyperref}
\usepackage{color}
\hypersetup{
  pdftitle= ,
  pdfsubject= ,
  pdfauthor=Dr. Mulugeta ,
  pdfmenubar=true,
  pdftoolbar=true,
  pdfpagemode={None},
  colorlinks=false,
  bookmarks=true,
  pdfkeywords= Water,
  }
\usepackage{graphics}
\usepackage{graphicx} 
\usepackage{latexsym,amssymb}
\title{\bf Unbinding transitions of   membranes and strings from single and double-well potential }
\author{Mesfin Asfaw\\ Department of Physics and  Graduate Institute of Biophysics\\
   National Central University,  Jhongli, 32001 Taiwan }
    
\begin{document}
\maketitle

\begin{abstract} 
We present a theory of unbinding transitions for membranes that interact via short and long receptor/ligand bonds.
 The detail of  unbinding behavior of the membranes is governed by the binding energies and concentrations of receptors and ligands. We investigate the unbinding behavior of these membranes with Monte Carlo simulations and via a comparison with strings.  We  derive the scaling laws for strings analytically. The exact analytic results  provide scaling estimate for membranes  in the vicinity of the critical point. 
\end{abstract}

\section{\bf Introduction}

Biological membranes consist of a multi-component lipid bilayer with different types of embedded or absorbed macromolecules \cite{h1,h2}. They perform a number of general functions in our cells and tissues. For instance, membranes separate cells and cell compartments. They also facilitate the transport of ions and macromolecules into and out of the cells. Some membrane proteins mediate interactions between membranes and participate in cell adhesion \cite{h2}. Since  membranes play a  vital role in biological processes, there are considerable experimental and theoretical interests [3-18].

Membranes undergo both lateral phase separation and unbinding transition. 
When two   membranes interact via   short-range  attractive potential, the attractive potential forces the membranes to bind.   Membranes also exhibit thermally excited shape fluctuations which compete with the molecular force potential. When  thermal fluctuations of the membranes are strong  enough, membranes undergo a transition from bound state to unbound state at a certain critical temperature $T_{u}$ and such transition is called unbinding transition.

The study of unbinding transitions of multicomponent membranes has received a significant attention [7,18-22]. In our recent theoretical work \cite{h21}, we presented a statistical-mechanical model of membranes that interact via two species of receptor/ligand bonds. Tracing out the receptor and ligand degrees of freedom  in the partition function leads to an effective double-well potential with effective depths  $U_{1}^{eff}$ and  $U_{2}^{eff}$. The critical point of lateral phase separation  was determined as a function of model parameters. We also  predicted the unbinding transition lines for membranes interacting with short and long receptor/ligand bonds by considering membranes that interact via a single-well potential.

In the present work,  instead of limiting the study of   unbinding transition of membranes  to membranes that interact with an effective single-well potential, we consider membranes that interact via an effective double-well potential. This will introduce some additional parameters to the model and, thereby, address a more general problem. We explore the unbinding transition of these membranes via exact analytic results of strings and by comparison with Monte Carlo simulation results.

 Some model systems like strings play a crucial role to study the unbinding transition of membranes.  Strings are lines governed by tension \cite{h24}.
  Functional renormalization group calculations show that  membranes  have similar scaling properties as strings in the vicinity of the critical potential depth \cite{h25}. Thus, it is  worth exploring the scaling behavior of strings. Qualitative similarities between phase diagrams for multi-component membranes and strings have been reported in the work of  \cite{h26}. In this paper we  derive the scaling law for unbinding critical potential depth of strings and suggest a deeper analogy between strings and membranes.

The rest of the  paper is organized as follows: In section II  we present the unbinding transition of strings that interact with square-well
  potential.
   We first give detailed calculation for strings that interact with single-well potential. We then  expose the scaling laws of strings that interact with double-well potential.  
 The strings scaling law is then  compared with Monte Carlo simulation results of membranes in section III. Section IV deals with     summary and conclusion.

\section{String model}

Strings are one dimensional lines where their shape fluctuations are governed by a finite tension \cite{h24}. We consider here two interacting
strings in two dimensional spaces. The conformation of strings can be  described by the local separation $l(x)$ perpendicular to the reference line where $x$ measures the distance along the reference line. The strings are, on average, parallel to this line.  The effective Hamiltonian of the model  
\begin{equation}
H \left\{ l \right\} =  \int_{0}^{L} 
\left[ {\sigma\over 2} \left( {dl\over dx} \right)^2 
+V^{'}(l) \right] dx
\label{1}
\end{equation}
contains the potential energy $V^{'}(l)$ and $\sigma$ which  denotes the effective tension of 
the strings.

 In the thermodynamic limit the statistical behavior of the model (1) can be explored by transfer matrix method which leads to 
 the Schr\"odinger-type equation \cite{h24,r1,r2}:
 \begin{equation}
-{T^{2}\over 2\sigma }{\partial^2 \psi_{k} \over \partial l^2} 
+ V^{'}(l) \psi_{k}(l) = E^{'}_{k} \psi_{k}(l).
\label{2}
\end{equation}
  Introducing  dimensionless variables $E_{k} = 2E_{k}^{'} / \sigma $, $z_{i} = l_{i} \sigma / T$ and $V(z_{i}) = 2V^{'}(l_{i}) / \sigma$, we rewrite Eq. (2) as 
 \begin{equation}
-{ \partial^2 \psi_{k} \over \partial z^2} 
+ V(z) \psi_{k}(z) = E_{k} \psi_{k}(z).
\label{2}
\end{equation}
 The parameters $E_{k}$ and $\psi_{k}$ denote the set of eigenvalues and the wave functions, respectively.
The set of eigenvalues $\left\{ E_{k} \right\}$ for the above equation is ordered in such a way that $E_{0} \le E_{1} \le E_{2} \ldots$.  The ground-state eigenvalue $E_{0}$ gives the free-energy density of the string, $f= E_{0}$, while the corresponding eigenvector $\psi_{0}(z)$ determines the probability distribution  $P(z)$. The probability distribution  $P(z)$ of  finding the string at distance $z$ from the reference line is given by
\begin{equation}
P(z)=  {| \psi_{0}(z)|^{2} \over
\int| \psi_{0}(z) |^{2} dz  \quad} .
\label{3}
\end{equation}
The mean and the first moment of the probability distribution is given by  
$ \langle z \rangle = \int z P(z) dz$
and
$\langle z^{2} \rangle = \int z^{2} P(z)dz$, respectively while the string roughness can be written as  $\xi_{\perp}= (\langle z^2 \rangle- \langle z \rangle^{2})^{1/2}$. On the other hand, in the thermodynamic limit,  the parallel correlation length $\xi_{||}$ can be expressed as $\xi_{||}=1/(E_{1} - E_{0})$. In the limit $E_{1} \to  E_{0}$,  the correlation length $\xi_{||}$ diverges which is the sign of a  continuous phase transition taking place in the system.

\subsection{Unbinding transition}
\subsubsection{Single-well potential}
From functional renormalization arguments, the scaling behavior of membranes interacting via single-well potential (as shown in Fig. 1) is similar to the scaling behavior of strings interacting via a single-well potential. Strings can be studied with analytical methods. First we explore the behavior of the critical potential depth  of strings as a function of the model parameters and finally compare with Monte Carlo results for membranes.

Consider strings interacting via a square-well potential 
\begin{equation}
V(z) =
\cases
{
\infty & for $z<0$\cr
-U & for  $z_{1}<z<z_{2}$\cr
0 & otherwise \cr
}
\label{potential}
\end{equation}
as shown in Fig. 1.
\begin{figure}
\begin{center}
\epsfig{file=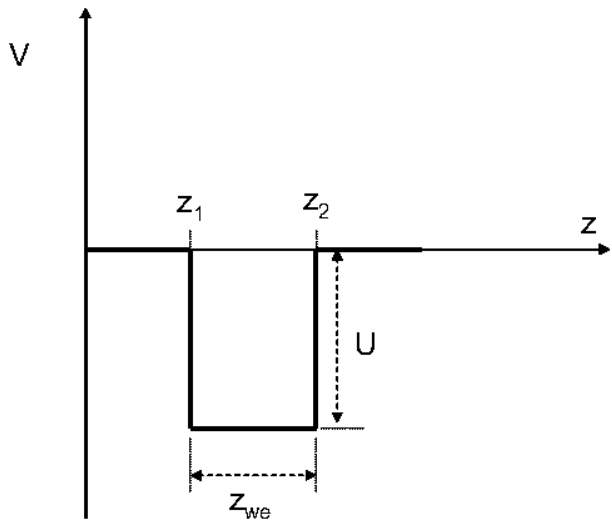,width=8cm}
\caption{The  potential $V$ versus $z$. The potential has one  square well  $U$  within the range $z_{2} - z_{1}=z_{we}$. }
\label{a4}
\end{center}
\end{figure}
The differential Eq. (3) can be easily solved  for the square-well potential (5) and 
the transfer-matrix eigenfunction for the  square-well potential (5)  has the
following form
\begin{equation}
\psi_{0}(z) = 
\cases
{
0 & for $z<0$\cr
A_{1} \exp (k z) - A_{1} \exp (-k z) & for $0<z<z_{1}$\cr
A_{2} \cos (\alpha z) + A_{3} \sin (\alpha z) & for $z_{1}<z<z_{2}$\cr
A_{4} \exp (-k z) & for $z>z_{2}$\cr
} .
\end{equation}
Here  $A_{1}$, $A_{2}$, $A_{3}$ and $A_{4}$  are coefficients which are independent of $z$ while the parameters $\alpha$ and $k$ are given by $ k =\sqrt{ -E_{0} } $ and $\alpha =\sqrt{E_{0}+ U}$. One should note that when $E_{0}<0$ the parameter $k$ takes real values and the wave function $\psi_{0}(z)$  decays as $z$ goes to infinity.  This implies that the probability distribution $P(z)$  also decays  with $z$. In this case the average string position $\langle z \rangle$ as well as the string roughness $\xi_{\perp}$ have  finite values.  When $E_{0}$ increases (but remains negative), the value of  $k$ gets smaller and the 'tail' of $\psi_{0}(z)$ lengthens. The probability distribution $P(z)$ becomes  broader and hence the average distance  $\langle z \rangle$ as well as the string roughness $\xi_{\perp}$ increase. At $E_{0} =0$ the unbinding transition takes place at which the distribution $P(z)$ becomes flat while $\langle z \rangle$ and $\xi_{\perp}$ diverge. 
\begin{figure}[h]
\epsfig{file=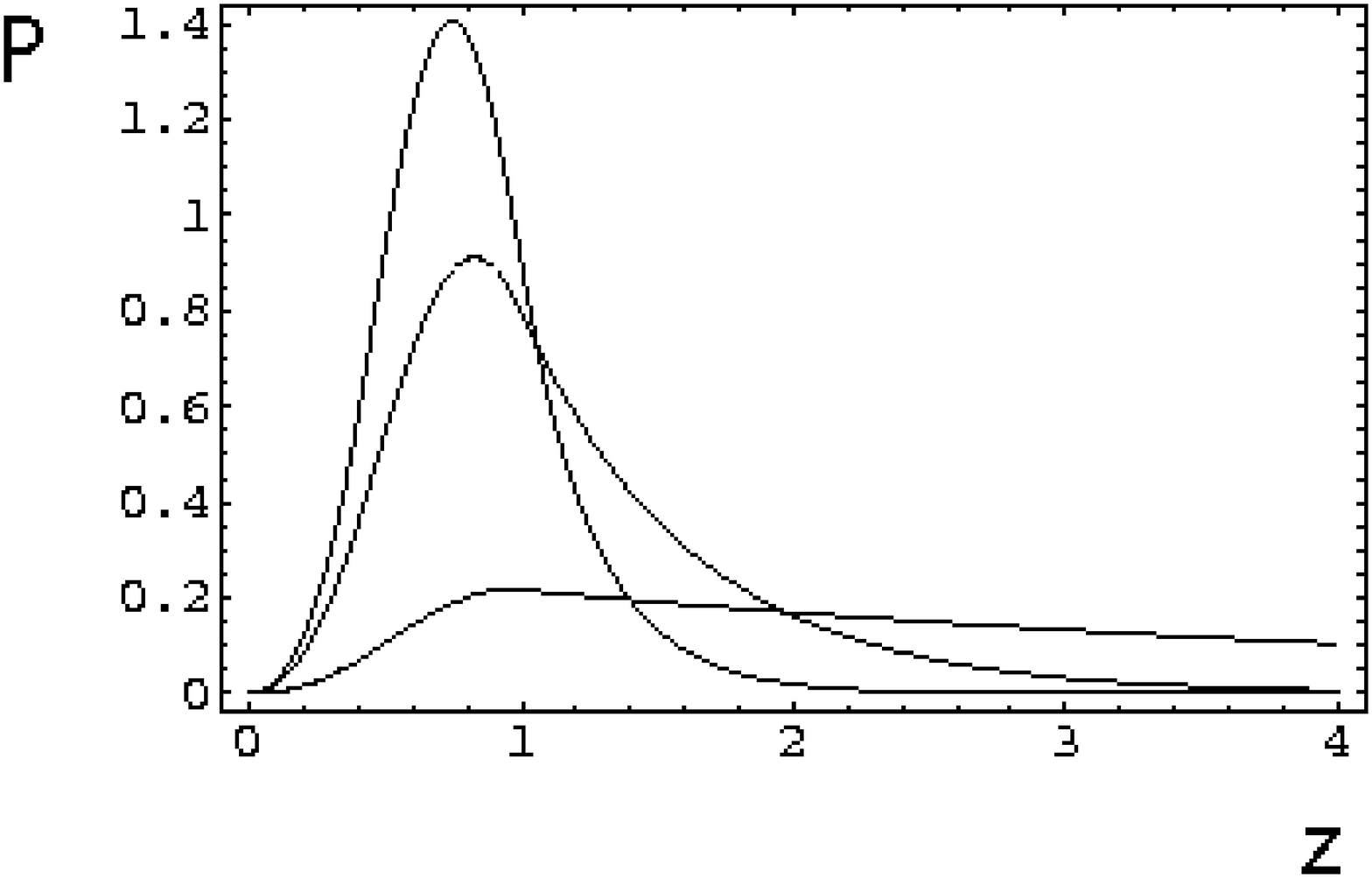,width=7cm}
\caption{Probability distribution of strings $P$ versus $ z$ for fixed   $z_{we}=0.6$ and  $z_{1}=0.4$. The potential depths  are fixed as   $U =10$,   $U=5$ and   $U =3$ from top to bottom. When the potential depth $U$ decreases the probability distribution becomes broader and flatter. }
\end{figure}

The wave function   $\psi_{0}(z)$ and its first derivative
$\partial_{z} \psi_{0}(z)$ should be continuous at $z=z_{1}$ and $z=z_{2}$. These requirements  lead to  four continuity conditions which finally guide to  a transcendental equation.  The transcendental equation in 
principle allows us to determine the free-energy density of the
string, $f=E_{0}$.  This transcendental equation is given by 
\begin{equation}
2k \exp{(2z_{1} k)}\alpha \eta_{1}-(-U+\exp{(2z_{1} k)}(-2k^2+U))\eta_{2}=0
 \end{equation}
 where $\eta_{1}=\cos(z_{we}\alpha)$ and $\eta_{2}=\sin(z_{we}\alpha)$. Here $z_{we}=z_{2}-z_{1}$. 
  
In the limit  $E_{0} \to 0$, the critical  potential depth $ U_{c}$ can be obtained from Eq. (7) as
\begin{equation}
\cos \left[ z_{we}  \sqrt{ U_{c}} \right] = z_{1}
\sqrt{ U_{c} } \sin \left[ z_{we}  \sqrt{ U_{c}}
\right] .
\label{9}
\end{equation}
When $z_{1} =0$, the above equation converges to a much simpler
expression, $\cos ( z_{2} \sqrt{ U_{c} } ) =0$, which implies
$U_{c} = {\pi^{2} \over 4 z_{2}^2}$.  For $z_{1} \neq 0$
equation (8) can be rearranged to
\begin{equation}
z_{1} \sqrt{ U_{c}}
\tan \left[ z_{we} \sqrt{ U_{c}} \right] =1 .
\label{10}
\end{equation}
It is worth to note that for the finite potential width
$z_{we}$, the  critical potential depth $U_{c}$ 
is different from zero.  It means that the unbinding transition takes place at a
finite temperature $T_{c}$ (and therefore it is often called a
non-trivial transition).  The strings thus are bound in the potential
well at low temperature $T<T_{c}$ (or when $U > U_{c}$)
and unbound from the wall at high temperature $T>T_{c}$ (when $U
< U_{c}$). 
\begin{figure} 
\begin{center}
\epsfig{file=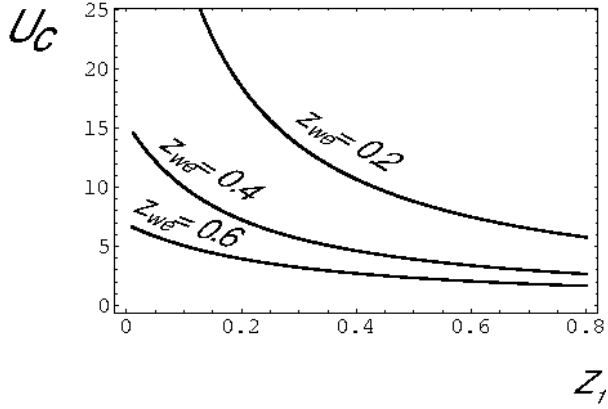,width=8cm}
\caption{ The critical potential depth ${U}_{c}$  as a function of the separation field  $z_{1}$. The  critical potential depth ${U}_{c}$ decreases as the parameters $z_{we}$ and $z_{1}$ increase. ${U}_{c} \to  0$  in the limit   $z_{we} \to \infty$ or $z_{1} \to \infty$ while ${U}_{c} \to \infty $  in the limit   $z_{we} \to 0$ or $z_{1} \to 0$. Here     
 the potential width is fixed as $z_{we}=0.2,~0.4$ and $0.6$. }
\label{ss5}
\end{center}
\end{figure}

 For  $U > U_{c}$, numerically we find how the probability distribution behaves as a function of $z$ as shown in Fig. 2. The figure clearly shows  that  the strings are strongly localized for the deep potential well. As the potential depth $U$ 
decreases, the probability distribution $P$ gets flatter.

The numerical solution to Eq. (9) gives us how  $U_{c}$ behaves as 
function of $z_{1}$ and $z_{we}$ as displayed in Fig. 3.  Figure 3 shows that
$U_{c}$ decreases monotonously as $z_{1}$ and $z_{we}$ increase.
 This effect can be easily understood.  When the
distance $z_{1}$ increases, the entropic repulsion between
strings and the hard wall become weaker. Thus the strings unbind at a
lower critical potential depth $U_{c}$.  When $z_{1}$ goes to infinity,
strings do not experience the presence of the wall and unbind at
$U_{c}=0$.   On the other hand when  $z_{we}$ increases, the entropic loss due to the confinement of strings  in the potential well decreases  and the strings unbind
at shallow critical point  $U_{c}$.

For fixed  $z_{we}=z_{2}-z_{1}$ and in the limit $z_{1}\to \infty$, the effect of the wall is negligible and  this corresponds to the case where strings  interact with symmetric square well potential which undergoes 
a delocalization transition at $U_{C}=0$. When  $z_{1}\to \infty$, Eq. (7) takes a simple form: $2k\alpha\eta_{1} = (-2k^{2}+U)\eta_{2}$. This equation can be  rewritten as 
\begin{eqnarray}
2\sqrt{-e}\sqrt{e+U} = (2e+u) \tan(z_{we}\sqrt{e+U}).
\end{eqnarray}
Using trigonometric identity  $\tan(z_{we}\sqrt{e+U})={2 \tan(0.5z_{we}\sqrt{e+U})\over(1-(\tan(0.5z_{we}\sqrt{e+U}))^2)}$ and  applying  this trigonometric property, one can rewrite Eq. (10) as 
\begin{eqnarray}
2\sqrt{-e}\sqrt{e+U}(\tan(0.5z_{we}\sqrt{e+U}))^2+  \nonumber \\
(4e+2v)\tan(0.5z_{we}\sqrt{e+U})-2\sqrt{-e}\sqrt{e+U}=0
\end{eqnarray}
Solving the quadratic equation (11) for $\tan(0.5z_{we}\sqrt{e+U})$, one gets
\begin{equation}
\sqrt{E_{0} + U } 
\tan \left( 0.5 z_{we} \sqrt{E_{0} + U } \right) 
= \sqrt{ -E_{0}} .
\label{4.21}
\end{equation}
One can easily notice that when $E_{0} =0$,   $U =0$.  This implies the
delocalization transition takes place when the potential depth
$U$ approaches to zero.  In order to explore the thermodynamic behavior
of the system in the vicinity of the transition point
($U=0$), we expand the transcendental equation (12) for a 
small dimensionless parameter $0< U \ll 1$. We  get a simple
expression for the ground-state eigenvalue $E_{0}$ which can be written as 
$
E_{0} \approx - {1 \over 16} U^{2} z_{we}^{2} .
$
One should note that the free energy density is given by  $f= E_{0}$ and therefore the
 free-energy density
of the string near to the transition point scales as 
 $f \sim - U^{2} z_{we}^{2}$.
From this simple scaling law, 
one can predict the scaling for the contact probability $P_{C}$ as $P_C \sim - U z_{we}^{2}$.

The transcendental equation (12) can be rederived  for strings interacting via symmetric single-well potential and 
the method of solving  such system  is
well known  \cite{dav} and we will not present it explicitly here. 
   Figure 4a
 shows the probability distribution $P(z)$ for string interacting with symmetric square-well potential of  width $z_{we}$.  When the potential well is deep, the
strings are strongly localized.  As the potential depth $U$ is
decreased, the probability distribution $P(z)$ gets flatter and
broader.  At $U=0$ the delocalization transition takes place.
We  also study the behavior of  the rescaled probability distribution $P(z/\xi_{\perp})$  in the vicinity of the critical point as  function of $z/\xi_{\perp}$. Figure 4b depicts that after rescaling all the rescaled probability distributions collapse into one scaling function. This reveals that near to the critical point the probability distribution exhibits the scaling form $P(z)=\xi_{\perp}^{-1}\Omega(z/\xi_{\perp})$. 
\begin{figure}[h]
\epsfig{file=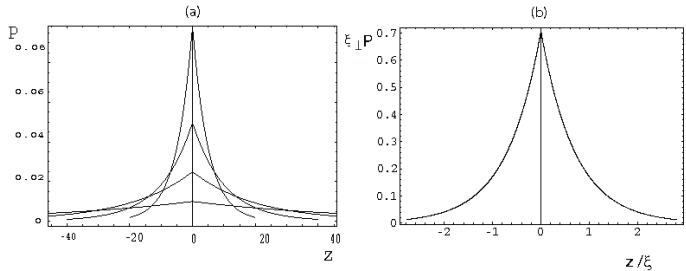,width=9cm}
\caption{(a)Probability  distribution $P$ versus  $z$ for  fixed  values $z_{we}=0.5$. The potential depth is fixed as  $U=0.4$, $U=0.2$, $U=0.1$ and $U=0.04$ from top to bottom. (b) Rescaled probability  distribution  $\xi_{\perp}P$  versus   $z/\xi_{\perp}$. After rescaling all the curves shown in Fig. 4a collapse onto one scaling function.
}
\end{figure}
\subsubsection{Double-well potential}
Consider membranes with short and long ligand/receptor bonds. The equilibrium phase behavior of such membranes is governed by an effective double-well potential. Functional renormalization group calculations reveal that membranes have the same scaling behavior as strings in the vicinity of the critical point. The unbinding critical potential depth for strings can be solved exactly using transfer matrix method. First we study the behavior of the critical potential depth  of strings that interact with double-well potential as shown in Fig. 5 and latter compare with Monte Carlo results for membranes  that interact via double-well potential ( see Fig. 5).

\begin{figure}
\begin{center}
\epsfig{file=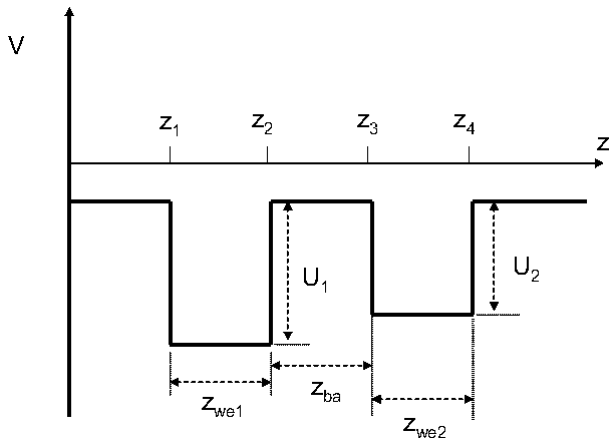,width=8cm}
\caption{ The potential $V$ versus $z$. The potential has two square wells of $U_{1}$ and $U_{2}$ within the ranges of $z_{we1}$ and $z_{we2}$, respectively. The parameter $z_{ba}$ separates the two  potential wells. }
\label{a5}
\end{center}
\end{figure}

Let us now consider strings interacting via double-well potential  as  shown in Fig. 5. The potential represents the hard wall located at $z=0$ and two potential wells of rectangular shape. The wave function which satisfies the differential equation (3) has the following form
\begin{equation}
\psi_{0}(z) =
\cases
{ 0&for $z<0$\cr
A_{1} \exp (k z) - A_{1} \exp (-k z) & for $0<z<z_{1}$\cr
A_{2} \cos ( \alpha_{1} z ) 
+ A_{3} \sin ( \alpha_{1} z ) & for $z_{1}<z<z_{2}$\cr
A_{4} \exp (kz) +A_{5} \exp (-k z) & for $z_{2}<z<z_{3}$\cr
A_{6} \cos (\alpha_{2} z) 
+ A_{7} \sin ( \alpha_{2} z ) & for $z_{3}<z<z_{4}$\cr
A_{8} \exp (-k z) & for $z>z_{4}$\cr
}.
\end{equation}
The transcendental equation which allows  determining  the smallest eigenvalue  is very complex in this case. 
Since we are interested in
finding the unbinding point, we take the limit $E_{0} \to 0$ and obtain the transcendental equation for $U_{1c}$ and $U_{2c}$ as
\begin{equation}
{\sqrt{U_{1C}}(z_{1}\sqrt{U_{1C}}k_{1}+(z_{1}+z_{ba})\sqrt{U_{2C}}k_{2})\over \sqrt{U_{1C}}k_{3}+(z_{1}z_{ba}U_{1C}-1)\sqrt{U_{2C}}k_{4}}=1
\end{equation}
where $ k_{1}= \cos{m_{2}}\sin{m_{1}}$, $k_{2}=\cos{m_{1}}\sin{m_{2}}$, $k_{3}=\cos{m_{1}}\cos{m_{2}}$ and  $k_{4}=\sin{m_{1}}\sin{m_{2}}$.
We have also introduced : $z_{we1} = z_{2}
- z_{1}$, $z_{we2} = z_{4} - z_{3}$, $z_{ba} = z_{3} - z_{2}$, $m_{1} =
z_{we1} \sqrt{U_{1c}}$ and $m_{2} = z_{we2} \sqrt{ U_{2c}}$.
\begin{figure} 
\epsfig{file=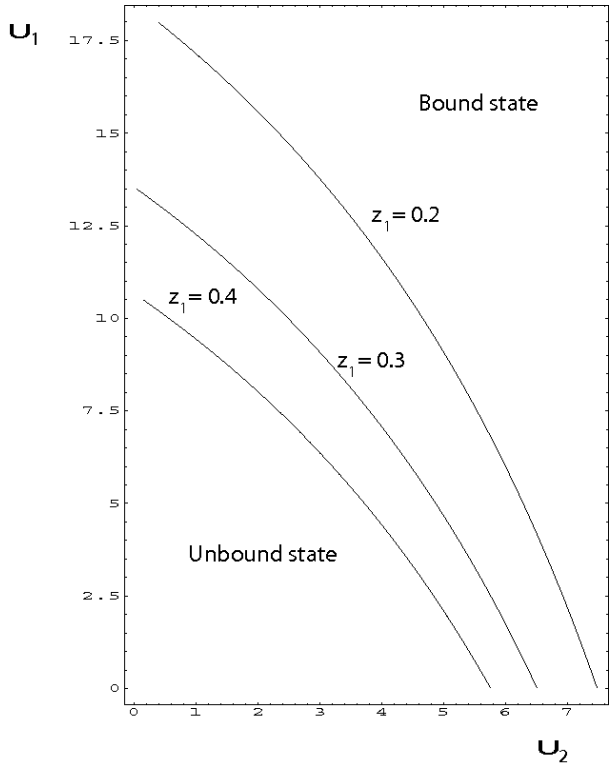,width=8cm}
\caption{ The phase diagram  of strings that interact via double-well potential. The strings are unbound for small values of   $U_{1}$ and  $U_{2}$. Here we take    $z_{we1}=z_{we2}=0.2$ and $z_{ba}=0.2$. As the parameter $z_{1}$ increases
 the phase boundary shifts down.}
\end{figure}
\begin{figure}
\epsfig{file=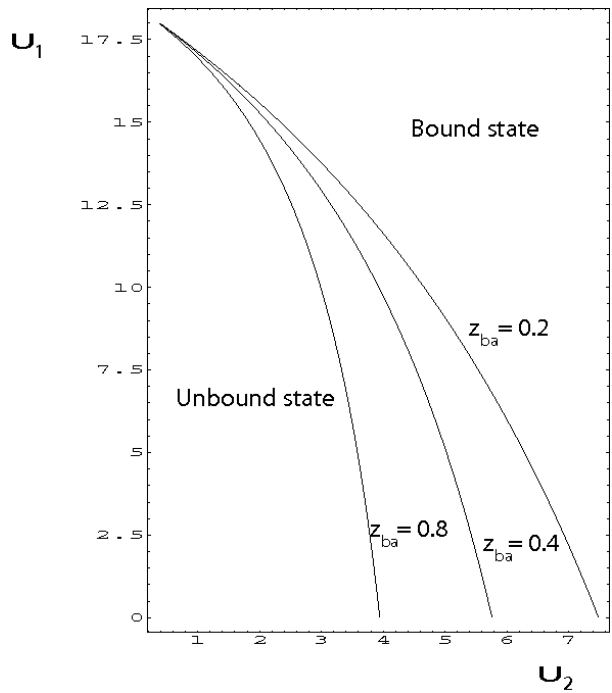,width=8cm}
\caption{ The phase diagram  of strings that interact via double-well potential. We take   $z_{we1}=z_{we2}=0.2$ and  $z_{1}=0.2$. As the parameter $z_{ba}$ increases
  the phase boundary shifts to the left.}
\end{figure}
One should note that in the limit $z_{3} \to \infty$, Eq. (14) converges to  Eq. ( 9) as one expects.

Equation (14) allows us to study how  the unbinding critical potential depths  $U_{1c}$ and $U_{2c}$ behave as  functions of the parameters characterizing the model. 
 Figure 6 shows the phase diagram in  $U_{1}$ and  $U_{2}$ parameter space for  $z_{ba} =0.2$ and
 $z_{we1} =z_{we2} = 0.2$.  As  demonstrated in the figure, as $z_{1}$ increases the phase boundary shifts down. One can note that when $z_{1}$ increases, the entropic repulsion of the strings with hard wall decreases. Therefore the strings unbind at  lower critical potential depth. 
  In the limit   $z_{1}$ goes to infinity, the strings do not feel the presence of hard wall. Hence in the limit 
  $z_{1} \to  \infty$,  $U_{1c} \to 0$ and  $U_{2c} \to 0$.  One should note that  even if $z_{we1}=z_{we2}$, the phase diagram 6 is  asymmetric due to the fact that  strings in the first potential well experience higher entropic repulsion from the hard wall than the strings that are confined in the second potential well. The effect of $z_{ba}$ on the phase diagram is also investigated.
 Figure 7 depicts  the phase diagram in the parameter space  $U_{1}$ and  $U_{2}$ for values $z_{we1}=z_{we2}=0.2$ and  $z_{1}=0.2$. As shown in the figure, as the parameter $z_{ba}$ increases
  the phase boundary shifts to the left.  It is important to note that since the values of $z_{1}$, $z_{we1}$ and $z_{we2}$ are fixed, when $z_{ba}$ increases only the position of the second potential well (well-two) shifts to the right. The critical potential depth $U_{1c}$ remains   unaffected while $z_{ba}$ increases. On the other hand, when  $z_{ba}$ increases, strings confined in the well-two experience a lesser entropic repulsion with the hard wall and due to this the strings unbind at lower critical potential depth $U_{2c}$.

\section{Comparison of membrane and string models}

\subsection{Unbinding from single-well potential}

Consider  membranes that  interact  with  receptor/ligand bonds of two different lengths. As presented in the work \cite{h21}, tracing out the receptor and ligand degrees of freedom  in the partition function leads to an effective double-well potential with potential wells of $U_{1}^{eff}$ and $U_{2}^{eff}$. In this section we consider membranes interacting via 
an effective single-well potential as shown in Fig. 1. 
We determine the unbinding critical potential depth  $U_{C}^{eff}$ with the Monte Carlo simulations. Within the Monte Carlo simulations, we consider the discretized Hamiltonian \cite{h21} and the effective single-well potential as shown in Fig. 1. The separation field $z_{i}$ of patch $i$ is shifted to another new value $z_{i}+\xi$. Here $\xi$ denotes a random number between $-1$ and $1$. We follow the standard Metropolis algorithm \cite{h28}. 
When the change  in configuration energy $\Delta H$ is negative, a local move is accepted;   when  $\Delta H$ is positive, the local move is accepted  with the probability  $exp[-\Delta H]$. In the simulations, membranes of size  $N=120X120$ patches are considered and to obtain a better statistics, the simulation is performed with up to $10^7$ attempted local moves per site $i$. In the vicinity of the critical point, both the autocorrelation time and correlation length diverge. Thus the simulation is performed for $U^{eff}>U^{eff}_{C}$.
 The critical point is obtained by measuring the contact probability $\left\langle P_{b}\right\rangle$ in the simulation. Here 
 $\left\langle P_{b}\right\rangle$ represents the expectation value for the fraction of bound membrane segments in the potential well. One should note that $\left\langle P_{b}\right\rangle$ is independent of the finite size of membranes \cite{h21}.  The critical point is determined by extrapolating of $\left\langle P_{b}\right\rangle$ as a function of $U^{eff}$ to the critical values 
$\left\langle P_{b}\right\rangle=0$. 

The plot of  $U_{C}^{eff}$ as a function of $z_{we}$ and $z_{1}$
is displayed in the Fig. 8. This result qualitatively agrees with the string result which  is shown in Fig. 3. The figure demonstrates that $U_{C}^{eff}$ decreases monotonously as $z_{we}$ and $z_{1}$ increase. When $z_{1}$ increases,  the steric repulsion of membranes with hard wall decreases. Therefore, membranes unbind at shallow critical potential depth. One should note that for membranes with one types of stickers, integrating  out stickers of degree of freedom leads to an effective potential in the partition function. Increasing the separation field $z_{1}$ corresponds to the increase in the length of the stickers. The result depicted in Fig. 8 shows that the unbinding transition takes place at lower critical potential depth (at higher temperature) when the length of the stickers increase. The depth of the critical point also  depends  on the width
 of the potential $z_{we}$. The analytic results of string (see Fig. 3) and the numerical results of membranes ( See Fig. 8) show that the critical potential depth  decreases as $z_{we}$ increases. When $z_{we}$ increases, the entropic loss due to confinement of strings in the potential well decreases and the strings or the membranes unbind at shallow critical point.  
\begin{figure}
\epsfig{file=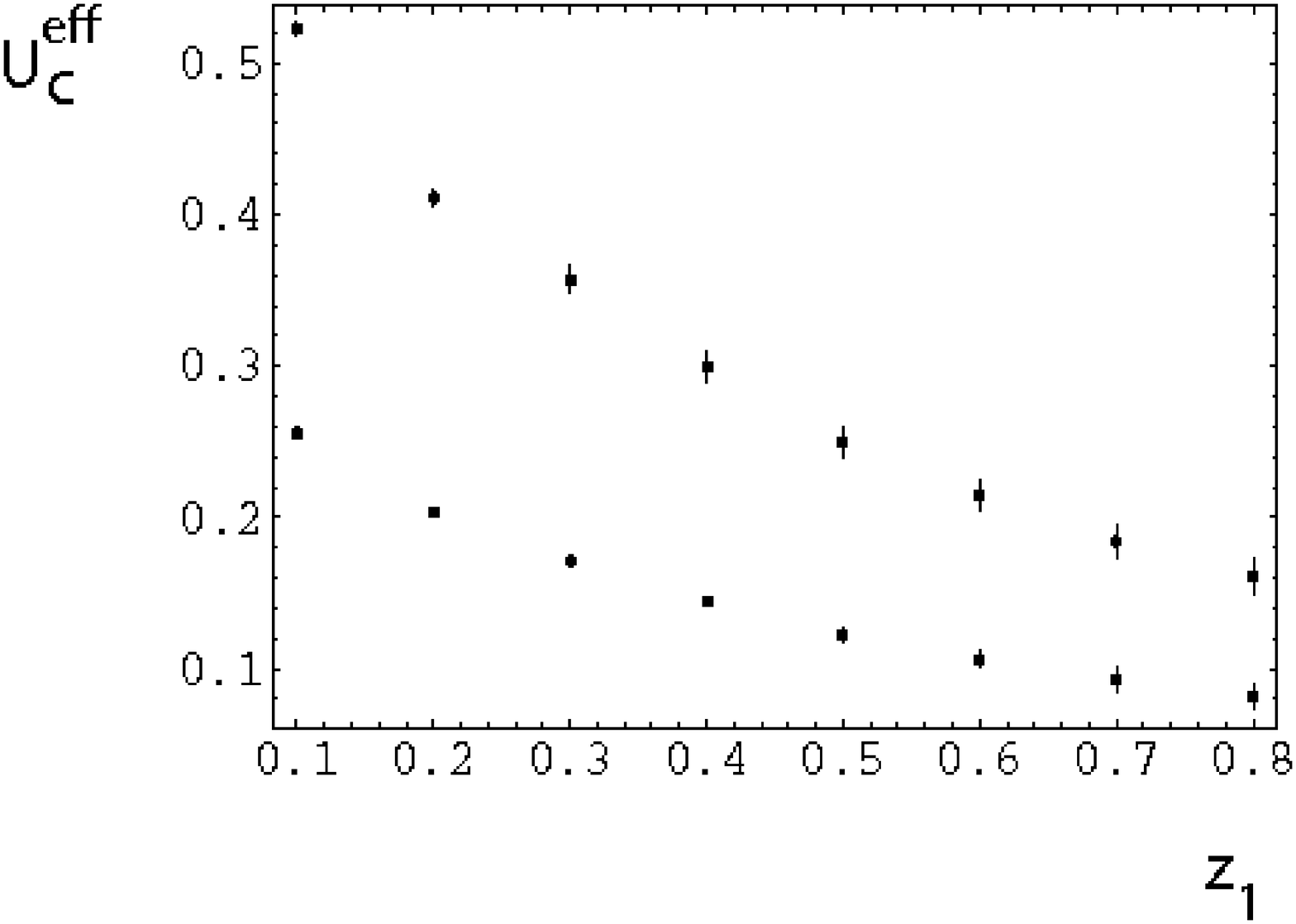,width=8cm}
\caption{Monte Carlo data  for critical potential depth $U_{C}^{eff}$ of unbinding from the single-well potential as a function of separation field   $z_{1}$ for  $z_{we}=0.2$ and $z_{we}=0.4$ from top to bottom. The critical potential depth $U_{C}^{eff}$  decreases as the parameters  $z_{1}$  and $z_{we}$ increase.}
\end{figure}

Functional renormalization indicates  that  the critical potential depth of 
membranes and strings have similar scaling properties  \cite{h24}.
Strings interacting via single-well potentials have scaling properties 
 $
\sqrt{\bar {U}_{c}}z_{1}\tan{\sqrt{\bar {U}_{c}}z_{we}}=1  
$
as discussed in the previous section.
Since membranes and strings have similar scaling properties, 
we  postulate  the relation 
\begin{equation}
\sqrt{{U}_{c}^{eff}}z_{1}\tan{\sqrt{{U}_{c}^{eff}}z_{we}}=C 
\end{equation}
to hold true for membranes. The constant $C$ can be obtained from data fitting.
\begin{figure}
\epsfig{file=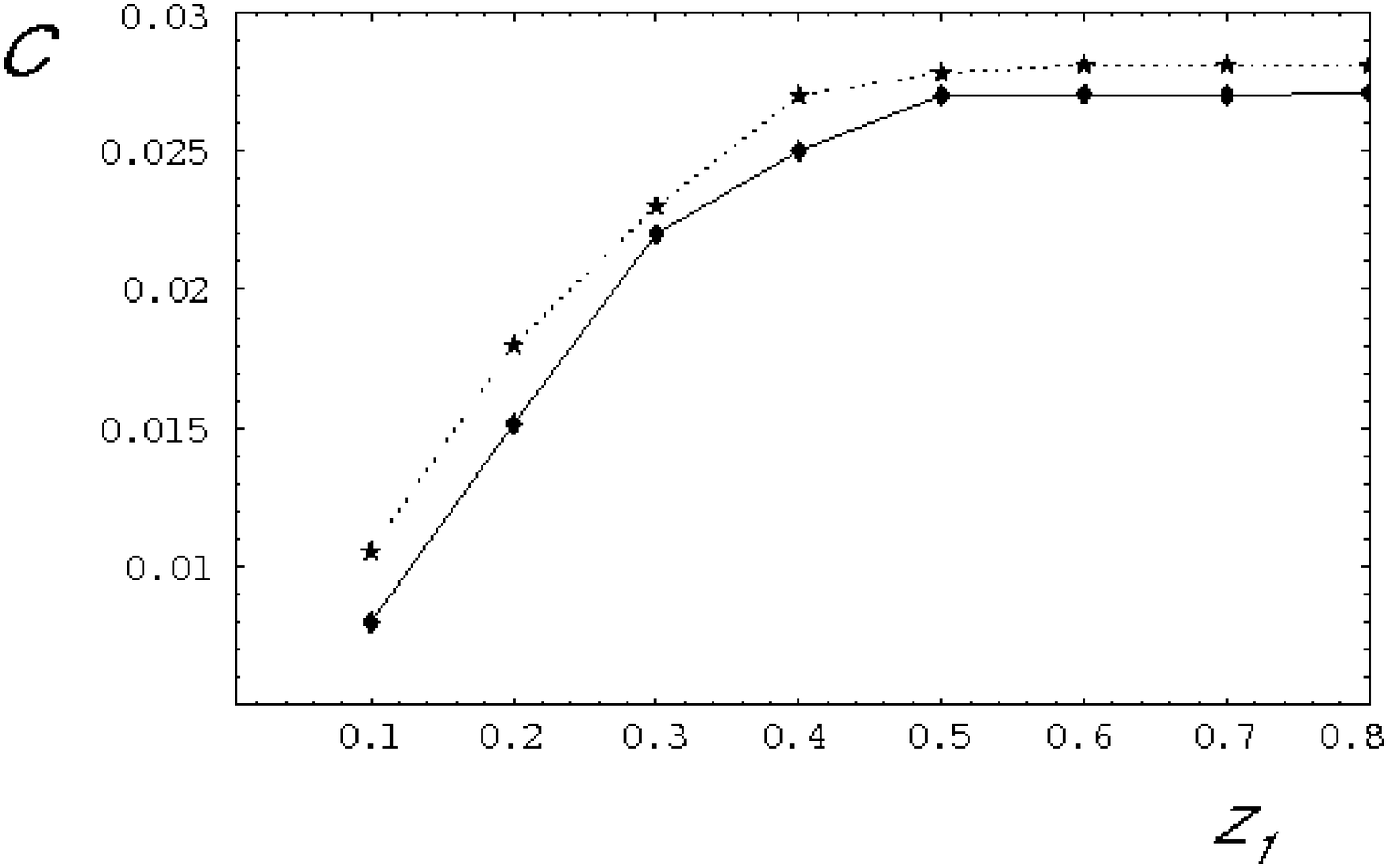,width=8cm}
\caption{The constant $C$ which represents the scaling law (15) for $z_{we}=0.4$ and $z_{we}=0.2$ from top to bottom. The value $C$ is obtained by substituting the Monte Carlo data (see Fig. 8) in Eq. (15).  The scaling law is valid for large values of  $z_{1}$  and  $z_{we}$.}
\end{figure}
There is a slight  difference between this work and the previous work \cite{h21}. In this work we don't consider the approximation $(z_{2}-z_{1})\sqrt{{{U}_{c}^{eff}}}<<1$. We substitute the corresponding values of ${{U}_{c}^{eff}}$, $z_{we}$ and $z_{1}$ ( see Fig. 8) in Eq. 15 and evaluate the constant $C$. Figure 9 shows how $C$ behaves as a function of $z_{1}$.
As indicated in  the figure,  when $z_{1}$ increases $C$ goes to a constant $ C=0.0281\pm 0.0024$. Here $C$ varies  for small values of $z_{1}$ since
we consider strings in the continuum limit while the membranes here  are 
descretized. In the discrete model, the continuum limit is reached for large values of 
$z_{1}$ and $z_{we}$. 

\subsection{Unbinding from  double-well potential}

Let us now consider 
 membranes that interact via short and long stickers. Tracing out stickers degree of freedom leads to membranes that interact with an effective double-well potential with potential wells of $U_{1}^{eff}$ and $U_{2}^{eff}$ \cite{h21}. Similar to the previous section, the unbinding critical potential depths $U_{1C}^{eff}$ and $U_{2C}^{eff}$ are determined in the simulation for different values of $z_{1}$, $z_{we}$ and $z_{ba}$. 

 Figure 10 reveals  the phase diagram in $U_{1}^{eff}$ and $U_{2}^{eff}$   parameter spaces for fixed $z_{2}-z_{1}= z_{we1}=0.3$, $z_{4}-z_{3}= z_{we1}=0.3$ and $z_{3}-z_{2}=z_{ba}=0.2$, $z_{1}=0.1,~0.2$ and $0.3$.
 The figure shows that as the parameter $z_{1}$ increases, the phase boundary shifts down.  It is important to note that increasing the separation field $z_{1}$ corresponds to the increase in the length of short and long stickers. 
The same figure demonstrates that as the length of short and long stickers increases, the critical potential depths $U_{1C}^{eff}$ and $U_{2C}^{eff}$  decrease. This result qualitatively agrees with the analytical result of string which is shown in Fig. 6.

We also  investigate the phase behavior of membranes  as a function of $z_{ba}$ for fixed $z_{1}$ and $z_{we}$. Our analysis demonstrates that as $z_{ba}$ increases, the phase boundary shifts to the left similar to the string result (see Fig. 7). Fixing  $z_{1}$ and $z_{we}$ implies fixing the length of short stickers while increasing  $z_{ba}$ corresponds to the increase in the length of long stickers. When one increases $z_{ba}$, the critical potential depth $U^{eff}_{1C}$ remains the same while the critical potential depth  $U^{eff}_{2C}$ decreases.
\begin{figure}
\epsfig{file=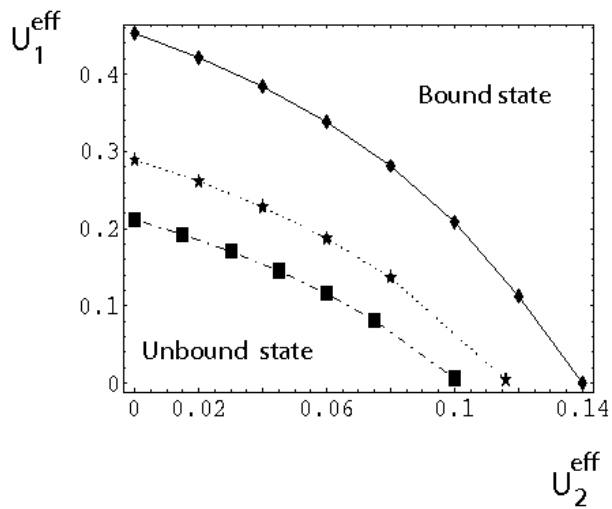,width=8cm}
\caption{Monte Carlo data  for critical potential depths of  $U_{1C}^{eff}$  and   $U_{2C}^{eff}$  for fixed $z_{2}-z_{1}= z_{we1}=0.3$, $z_{4}-z_{3}= z_{we1}=0.3$ and $z_{3}-z_{2}=z_{ba}=0.2$, $z_{1}=0.1,~0.2$ and $0.3$ from right to left.  $U_{1C}^{eff}$   and   $U_{2C}^{eff}$  decrease as $z_{1}$  increases. This result qualitatively  agrees with the result displayed in Fig. 6.}
\end{figure}

In addition to unbinding transitions, 
membranes also undergo lateral phase separation. When $z_{1} \ne 0$, the unbinding critical point and the critical point for the lateral phase separation are always detangled. The unbinding transition in this case is second  order while the lateral phase transition is first order. In the case of unbinding transition,  we can compare  membranes and strings either qualitatively or quantitatively as both exhibit continuous unbinding transitions. For strings  interacting via double-well potential, the transcendental  equation  for  $U_{1C}$ and $U_{2C}$ is given in Eq. (14). 
Because  membranes and strings have similar scaling properties, 
we  postulate  the relation 
\begin{equation}
{\sqrt{U_{1C}^{eff}}(z_{1}\sqrt{U_{1C}^{eff}}k_{1}+(z_{1}+z_{ba})\sqrt{U_{2C}^{eff}}k_{2})\over \sqrt{U_{1C}^{eff}}k_{3}+(z_{1}z_{ba}U_{1C}^{eff}-1)\sqrt{U_{2C}^{eff}}k_{4}}=C
\end{equation}
to hold true for membranes where
 $ k_{1}= \cos{m_{2}}\sin{m_{1}}$, $k_{2}=\cos{m_{1}}\sin{m_{2}}$, $k_{3}=\cos{m_{1}}\cos{m_{2}}$ and  $k_{4}=\sin{m_{1}}\sin{m_{2}}$.
 Here, $z_{we1} = z_{2}
- z_{1}$, $z_{we2} = z_{4} - z_{1}$, $z_{ba} = z_{3} - z_{2}$, $m_{1} =
z_{we1} \sqrt{U_{1c}^{eff}}$, $m_{2} = z_{we2} \sqrt{ U_{2c}^{eff}}$.
The constant $C$ is obtained from data fitting. We substitute the Monte Carlo data   $U_{1C}^{eff}$, $U_{2C}^{eff}$, $z_{we}$, $z_{ba}$
and $z_{1}$ in Eq. (16) and evaluate the constant $C$ as a function of $z_{1}$.
Similar to the previous section, the constant $C$ saturates to a certain constant $C$ for large values of $z_{1}$. Our analysis shows that at a given value of $z_{1}$, the constant $C$ do not vary significantly. However significant change in $C$ is observed as $z_{1}$ increases.  The numerical analysis  indicates that the constant $C$ saturates to a constant $C=0.026\pm 0.003$ as $z_{1}$ increases. Using Eq. (16) one can construct a  complete phase diagram for unbinding transition of membranes that interact via two species of receptor/ligand bonds. In the limit
$z_{1}\to \infty$  or $z_{3}\to \infty$, Eq. (16) converges to Eq. (15). This indicates that membranes interacting with single and double-well potentials should have the same constant $C$.

\section{Summary and conclusions}

In this article, we consider strings interacting via a single-well potential. The behavior of the critical potential depth $U_{c}$  of unbinding from a single-well as a function of model parameters is explored analytically. The critical potential depth  $U_{c}$  decreases when  $z_{1}$ and  $z_{we}$ increase.  In  the limit   $z_{we} \to \infty$ or $z_{ba} \to \infty$, ${U}_{c} \to  0$   while ${U}_{c} \to \infty $  in the limit   $z_{we} \to 0$ or $z_{ba} \to 0$.

For strings interacting with double-well potential, the  behavior of the critical points $U_{1c}$ and $U_{2c}$ is studied analytically. The critical potential depths  $U_{1c}$ and $U_{2c}$ are   functions  of $z_{we1}$, $z_{we2}$, $z_{1}$ and $z_{ba}$. $U_{1c}$ and $U_{2c}$ decrease as $z_{we1}$, $z_{we2}$, $z_{1}$ and $z_{ba}$ increase.  For fixed $z_{ba}$, in the limit $z_{we1}=z_{we2}=z_{we} \to  \infty$ or  $z_{1} \to \infty $,  $U_{1c}$ and $U_{2c}$ go to zero. When $z_{we1}=z_{we2}=z_{we}$  and  $z_{1} $ are fixed, the critical potential depth $U_{2c} \to 0$ as  $z_{ba} \to \infty $ while 
 $U_{2c} \to \infty$ as  $z_{ba} \to 0 $.
 
 The Monte Carlo simulation results show that the critical point  ${ {U}_{c}^{eff}}$ for membranes interacting via single-well potential decreases as $z_{1}$ and  $z_{we}$ increase similar to the result for strings that interact with single-well potential. On the other hand, the Monte Carlo studying
 for membranes in an effective double-well potential shows that the critical points  ${{U}_{1c}^{eff}}$ and ${{U}_{2c}^{eff}}$ decrease as the parameters 
 $z_{1}$, $z_{we}$ and $z_{ba}$ increase. We compare the Monte Carlo data of membranes with string analytic result.
 From scaling property of strings and Monte Carlo simulations, we find a new scaling behavior for membranes interacting via single-well and double-well potentials.

  \section*{\bf Acknowledgement}
It is my pleasure to thank Prof. R. Lipowsky,  T. R. Weikl and  B. Rozycki for interesting discussions during my stay at Max Planck institute of Colloids and Interfaces Potsdam, Germany. I would like  also to thank  Hsuan-Yi Chen and Mulugeta Bekele for stimulating discussions.

\end{document}